\pdfoutput=1

\documentclass[12pt,a4paper]{article}

\usepackage{ifthen} 
\usepackage{lmodern}
\usepackage{graphicx}

\newboolean{pdflatex}
\setboolean{pdflatex}{true} 

\newboolean{articletitles}
\setboolean{articletitles}{true} 

\newboolean{uprightparticles}
\setboolean{uprightparticles}{false}

\newboolean{inbibliography}
\setboolean{inbibliography}{false}

\usepackage{microtype}
\usepackage{lineno}  % for line numbering during review
\usepackage{xspace} % To avoid problems with missing or double spaces after
\usepackage{caption} %these three command get the figure and table captions automatically small

\usepackage{graphicx}  % to include figures (can also use other packages)
\usepackage{color}
\usepackage{colortbl}
\graphicspath{{./figs/}} % Make Latex search fig subdir for figures

\usepackage{amsmath} % Adds a large collection of math symbols
\usepackage{amssymb}
\usepackage{amsfonts}
\usepackage{upgreek} % Adds in support for greek letters in roman typeset

\newcommand*\patchAmsMathEnvironmentForLineno[1]{%
\expandafter\let\csname old#1\expandafter\endcsname\csname #1\endcsname
\expandafter\let\csname oldend#1\expandafter\endcsname\csname
end#1\endcsname
 \renewenvironment{#1}%
   {\linenomath\csname old#1\endcsname}%
   {\csname oldend#1\endcsname\endlinenomath}%
}
\newcommand*\patchBothAmsMathEnvironmentsForLineno[1]{%
  \patchAmsMathEnvironmentForLineno{#1}%
  \patchAmsMathEnvironmentForLineno{#1*}%
}
\AtBeginDocument{%
\patchBothAmsMathEnvironmentsForLineno{equation}%
\patchBothAmsMathEnvironmentsForLineno{align}%
\patchBothAmsMathEnvironmentsForLineno{flalign}%
\patchBothAmsMathEnvironmentsForLineno{alignat}%
\patchBothAmsMathEnvironmentsForLineno{gather}%
\patchBothAmsMathEnvironmentsForLineno{multline}%
\patchBothAmsMathEnvironmentsForLineno{eqnarray}%
}

\usepackage{hyperref}    % Hyperlinks in references
\usepackage[all]{hypcap} % Internal hyperlinks to floats.

\def\lhcb {\mbox{LHCb}\xspace}

\def\MagUp {\mbox{\em Mag\kern -0.05em Up}\xspace}

\ifthenelse{\boolean{uprightparticles}}%
{

 \def\PDelta      {\ensuremath{\Delta}\xspace}                 
 \def\PXi      {\ensuremath{\Xi}\xspace}                 
 \def\PLambda      {\ensuremath{\Lambda}\xspace}                 
 \def\PSigma      {\ensuremath{\Sigma}\xspace}                 
 \def\POmega      {\ensuremath{\Omega}\xspace}                 
 \def\PUpsilon      {\ensuremath{\Upsilon}\xspace}                 
 
 \def\PB      {\ensuremath{\mathrm{B}}\xspace}                 
                  
 \def\PD      {\ensuremath{\mathrm{D}}\xspace}

 \def\PK      {\ensuremath{\mathrm{K}}\xspace}

 \def\Pi      {\ensuremath{\mathrm{i}}\xspace}

 \def\Pt      {\ensuremath{\mathrm{t}}\xspace}

}
{

 \mathchardef\PDelta="7101
 \mathchardef\PXi="7104
 \mathchardef\PLambda="7103
 \mathchardef\PSigma="7106
 \mathchardef\POmega="710A
 \mathchardef\PUpsilon="7107
                  
 \def\PB      {\ensuremath{B}\xspace}                 
                  
 \def\PD      {\ensuremath{D}\xspace}

 \def\PK      {\ensuremath{K}\xspace}

 \def\Pi      {\ensuremath{i}\xspace}

 \def\Pt      {\ensuremath{t}\xspace}

}

\makeatletter
\ifcase \@ptsize \relax% 10pt
  \newcommand{\miniscule}{\@setfontsize\miniscule{4}{5}}% \tiny: 5/6
\or% 11pt
  \newcommand{\miniscule}{\@setfontsize\miniscule{5}{6}}% \tiny: 6/7
\or% 12pt
  \newcommand{\miniscule}{\@setfontsize\miniscule{5}{6}}% \tiny: 6/7
\fi
\makeatother

\DeclareRobustCommand{\optbar}[1]{\shortstack{{\miniscule (\rule[.5ex]{1.25em}{.18mm})}
  \\ [-.7ex] $#1$}}

\def\tquark    {{\ensuremath{\Pt}}\xspace}
\def\tquarkbar {{\ensuremath{\overline \tquark}}\xspace}
\def\ttbar     {{\ensuremath{\tquark\tquarkbar}}\xspace}

  \def\Kbar    {{\kern 0.2em\overline{\kern -0.2em \PK}{}}\xspace}

\def\KorKbar    {\kern 0.18em\optbar{\kern -0.18em K}{}\xspace}

  \def\Dbar    {{\kern 0.2em\overline{\kern -0.2em \PD}{}}\xspace}

\def\DorDbar    {\kern 0.18em\optbar{\kern -0.18em D}{}\xspace}

\def\Bbar    {{\ensuremath{\kern 0.18em\overline{\kern -0.18em \PB}{}}}\xspace}

\def\BorBbar    {\kern 0.18em\optbar{\kern -0.18em B}{}\xspace}

  \def\Y#1S{\ensuremath{\PUpsilon{(#1S)}}\xspace}% no space before {...}!

\def\Lbar        {{\ensuremath{\kern 0.1em\overline{\kern -0.1em\PLambda}}}\xspace}
\def\LorLbar    {\kern 0.18em\optbar{\kern -0.18em \PLambda}{}\xspace}

\newcommand{\decay}[2]{\ensuremath{#1\!\to #2}\xspace}         % {\Pa}{\Pb \Pc}

\def\to                 {\ensuremath{\rightarrow}\xspace}

\def\ssqtw   {{\ensuremath{\sin^{2}\!\theta_{\mathrm{W}}}}\xspace}

\def\ssqtwef {{\ensuremath{{\sin}^{2}\theta_{\mathrm{W}}^{\mathrm{eff}}}}\xspace}

\newcommand{\as}{{\ensuremath{\alpha_s}}\xspace}

\def\AT#1     {\ensuremath{A_{\mathrm{T}}^{#1}}\xspace}           % 2

\def\C#1      {\ensuremath{\mathcal{C}_{#1}}\xspace}                       % 9
\def\Cp#1     {\ensuremath{\mathcal{C}_{#1}^{'}}\xspace}                    % 7
\def\Ceff#1   {\ensuremath{\mathcal{C}_{#1}^{\mathrm{(eff)}}}\xspace}        % 9  
\def\Cpeff#1  {\ensuremath{\mathcal{C}_{#1}^{'\mathrm{(eff)}}}\xspace}       % 7
\def\Ope#1    {\ensuremath{\mathcal{O}_{#1}}\xspace}                       % 2
\def\Opep#1   {\ensuremath{\mathcal{O}_{#1}^{'}}\xspace}                    % 7

\newcommand{\tev}{\ifthenelse{\boolean{inbibliography}}{\ensuremath{~T\kern -0.05em eV}\xspace}{\ensuremath{\mathrm{\,Te\kern -0.1em V}}}\xspace}
\newcommand{\gev}{\ensuremath{\mathrm{\,Ge\kern -0.1em V}}\xspace}
\newcommand{\mev}{\ensuremath{\mathrm{\,Me\kern -0.1em V}}\xspace}
\newcommand{\kev}{\ensuremath{\mathrm{\,ke\kern -0.1em V}}\xspace}
\newcommand{\ev}{\ensuremath{\mathrm{\,e\kern -0.1em V}}\xspace}
\newcommand{\gevc}{\ensuremath{{\mathrm{\,Ge\kern -0.1em V\!/}c}}\xspace}
\newcommand{\mevc}{\ensuremath{{\mathrm{\,Me\kern -0.1em V\!/}c}}\xspace}
\newcommand{\gevcc}{\ensuremath{{\mathrm{\,Ge\kern -0.1em V\!/}c^2}}\xspace}
\newcommand{\gevgevcccc}{\ensuremath{{\mathrm{\,Ge\kern -0.1em V^2\!/}c^4}}\xspace}
\newcommand{\mevcc}{\ensuremath{{\mathrm{\,Me\kern -0.1em V\!/}c^2}}\xspace}

\def\mum  {\ensuremath{{\,\upmu\rm m}}\xspace}

\def\fb   {\ensuremath{\mbox{\,fb}}\xspace}

\newcommand{\chisq}{\ensuremath{\chi^2}\xspace}
\newcommand{\chisqndf}{\ensuremath{\chi^2/\mathrm{ndf}}\xspace}

\def\deriv {\ensuremath{\mathrm{d}}}

\def\gsim{{~\raise.15em\hbox{$>$}\kern-.85em
          \lower.35em\hbox{$\sim$}~}\xspace}
\def\lsim{{~\raise.15em\hbox{$<$}\kern-.85em
          \lower.35em\hbox{$\sim$}~}\xspace}

\def\sqs   {\ensuremath{\protect\sqrt{s}}\xspace}

\def\ptot       {\mbox{$p$}\xspace}
\def\pt         {\mbox{$p_{\rm T}$}\xspace}

\def\evtgen     {\mbox{\textsc{EvtGen}}\xspace}
\def\fewz       {\mbox{\textsc{Fewz}}\xspace}

\def\geant      {\mbox{\textsc{Geant4}}\xspace}

\def\herwig     {\mbox{\textsc{Herwig}}\xspace}

\def\photos     {\mbox{\textsc{Photos}}\xspace}
\def\powheg     {\mbox{\textsc{Powheg}}\xspace}
\def\pythia     {\mbox{\textsc{Pythia}}\xspace}

\def\tell1  {TELL1\xspace}
\def\ukl1   {UKL1\xspace}

\usepackage{cite} % Allows for ranges in citations
\usepackage{mciteplus}

\usepackage{longtable}
\usepackage{multirow}
\usepackage{lscape}

\newcommand\powhegbox{\textsc{\mbox{Powheg-Box}}}
\newcommand\qqZymm{$q\bar{q} \rightarrow Z \rightarrow \mu^{+}\mu^{-}$}

\newcommand\zymm{$Z \rightarrow \mu^{+}\mu^{-}$}
\newcommand\ztautau{$Z \rightarrow \tau^{+} \tau^{-}$}
\newcommand\zy{\ensuremath{Z/\gamma^{\ast}}}
\newcommand\afb{\ensuremath{A_\text{FB}}}
\newcommand\afbm{\ensuremath{A_\text{FB}(m_{\mu\mu})}}
\newcommand\afbt{\ensuremath{A_\text{FB}^{\text{pred}}}}

\newcommand\ww{\ensuremath{\decay{W^{+}W^{-}}{\mu^{+}\nu_{\mu}\mu^{-} \bar{\nu_{\mu}}} } }

\newcommand\finalresulta{\ensuremath{\ssqtwef = 0.23219 \pm 0.00148}}
\newcommand\finalresultb{\ensuremath{\ssqtwef = 0.23074 \pm 0.00123}}

\newcommand\finalresultcombfull{\ensuremath{\ssqtwef = 0.23142 \pm 0.00073 \pm 0.00052 \pm 0.00056 }}

\newcommand\numf{\ensuremath{N_\text{F}}}
\newcommand\numb{\ensuremath{N_\text{B}}}

\begin{document}

\renewcommand{\thefootnote}{\fnsymbol{footnote}}
\setcounter{footnote}{1}

% $Id: title-LHCb-PAPER.tex 67452 2015-02-10 11:22:35Z roldeman $
% ===============================================================================
% Purpose: LHCb-PAPER journal paper title page template
% Author: 
% Created on: 2010-09-25
% ===============================================================================

%%%%%%%%%%%%%%%%%%%%%%%%%
%%%%%  TITLE PAGE  %%%%%%
%%%%%%%%%%%%%%%%%%%%%%%%%
\begin{titlepage}
\pagenumbering{roman}

% Header ---------------------------------------------------
\vspace*{-1.5cm}
\centerline{\large EUROPEAN ORGANIZATION FOR NUCLEAR RESEARCH (CERN)}
\vspace*{1.5cm}
\noindent
\begin{tabular*}{\linewidth}{lc@{\extracolsep{\fill}}r@{\extracolsep{0pt}}}
\ifthenelse{\boolean{pdflatex}}% Logo format choice
{\vspace*{-2.7cm}\mbox{\!\!\!\includegraphics[width=.14\textwidth]{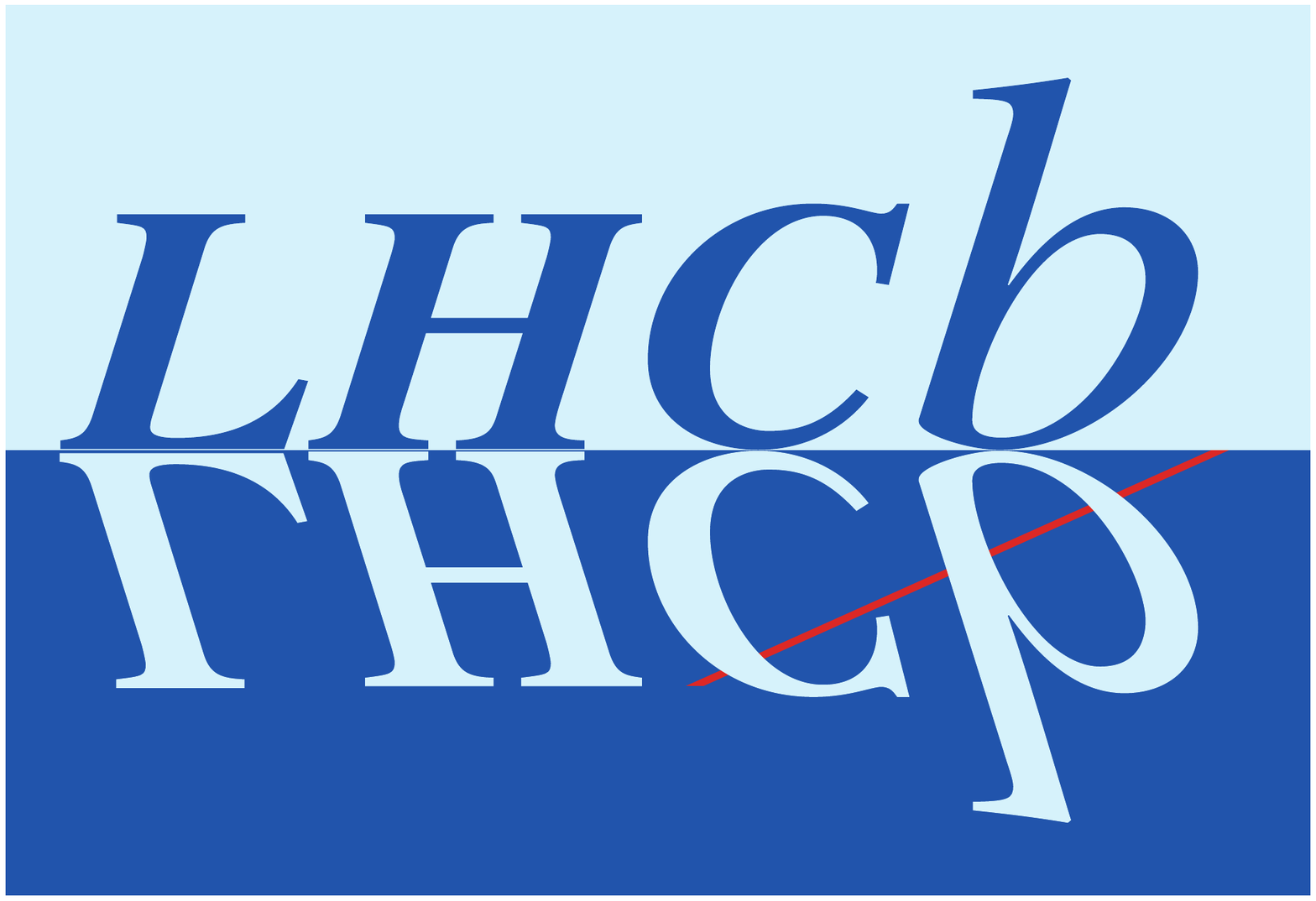}} & &}%
{\vspace*{-1.2cm}\mbox{\!\!\!\includegraphics[width=.12\textwidth]{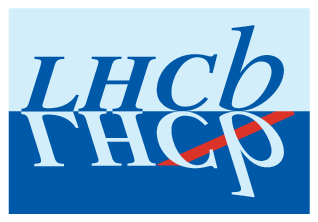}} & &}%
\\
 & & CERN-PH-EP-2015-250 \\  % ID 
 & & LHCb-PAPER-2015-039 \\  % ID 
 %& & \today \\ 
 & & 24 September 2015 \\ % Date - Can also hardwire e.g.: 23 March 2010
 & & \\
% not in paper \hline
\end{tabular*}

\vspace*{2.0cm}

% Title --------------------------------------------------
{\bf\boldmath\huge
\begin{center}
  Measurement of the forward-backward asymmetry in $Z/\gamma^{\ast} \rightarrow \mu^{+}\mu^{-}$ decays and determination of the effective weak mixing angle
\end{center}
}

%\vspace*{2.0cm}

% Authors -------------------------------------------------
\begin{center}
%In the footnote, replace 'paper' by 'letter' in case of submission to PRL or PLB 
The LHCb collaboration\footnote{Authors are listed at the end of this paper.}
\end{center}
\vspace{\fill}

% Abstract -----------------------------------------------
\begin{abstract}
  \noindent
The forward-backward charge asymmetry for the process $q\bar{q} \rightarrow Z/\gamma^{\ast} \rightarrow \mu^{+}\mu^{-}$ is measured as a function of the invariant mass of the dimuon system.
Measurements are performed using proton proton collision data collected with the LHCb detector at $\sqs = 7$ and 8\tev, corresponding to integrated luminosities of $1\fb^{-1}$ and $2\fb^{-1}$ respectively. 
Within the Standard Model the results constrain the effective electroweak mixing angle to be
$$\finalresultcombfull,$$
where the first uncertainty is statistical, the second systematic and the third theoretical.
This result is in agreement with the current world average, 
and is one of the most precise determinations at hadron colliders to date.
\end{abstract}

\vspace*{2.0cm}

\begin{center}
  %Submitted to JHEP 
  Published in JHEP 1511(2015) 190 
\end{center}

\vspace{\fill}

{\footnotesize 
\centerline{\copyright~CERN on behalf of the \lhcb collaboration, licence \href{http://creativecommons.org/licenses/by/4.0/}{CC-BY-4.0}.}}
\vspace*{2mm}

\end{titlepage}

%%%%%%%%%%%%%%%%%%%%%%%%%%%%%%%%
%%%%%  EOD OF TITLE PAGE  %%%%%%
%%%%%%%%%%%%%%%%%%%%%%%%%%%%%%%%

%  empty page follows the title page ----
\newpage
\setcounter{page}{2}
\mbox{~}
%\newpage

\cleardoublepage

\renewcommand{\thefootnote}{\arabic{footnote}}
\setcounter{footnote}{0}

\pagestyle{plain} 
\setcounter{page}{1}
\pagenumbering{arabic}

\section{Introduction}
\label{sec:Introduction}

In the Standard Model (SM), the $Z$ boson couplings differ for left- and right-handed fermions. 
The difference leads to an asymmetry in the angular distribution of positively and negatively charged leptons produced in $Z$ boson decays. 
This asymmetry depends on the weak mixing angle ($\theta_W$) between the neutral states associated to the U(1) and SU(2) gauge groups,
i.e. the relative coupling strengths between the photon and the $Z$ boson. 
In order to compare directly with previous experimental determinations, a scheme is adopted in which the higher order corrections to the $Z$ boson couplings are absorbed in effective couplings. 
The resulting effective parameter \ssqtwef is defined as a function of the ratio of the vector and the axial-vector effective couplings of the $Z$ boson to the fermions involved~\cite{ALEPH:2005ab}, 
and is proportional to \ssqtw.

Defining $\theta^{\ast}$ as the polar angle of the negatively charged lepton in the Collins-Soper~\cite{PhysRevD.16.2219} frame, in which the
direction of the $z$-axis is aligned with the difference of the incoming proton momentum vectors in the dimuon rest frame, 
the differential cross section in the SM at leading order is
\begin{equation*}
  \label{eq:stwafb}
  \frac{\deriv \sigma}{\deriv \cos \theta^{\ast}} =     A (1+\cos^{2}\theta^{\ast}) + B \cos \theta^{\ast}.
  \end{equation*}
\noindent
Here $A$ and $B$ are coefficients that depend on the dimuon invariant mass, mainly because of interference between $Z$ and $\gamma^{\ast}$ contributions, the colour charge of the quarks and the vector and axial-vector couplings. 
The parameter $B$ is a function of \ssqtw and is proportional to the forward-backward asymmetry \afb, which is given by
\begin{equation*}
  \label{eq:afb}
  \afb \equiv \frac{\numf - \numb}{\numf + \numb},
  \end{equation*}
\noindent
where 
\numf~represents the number of forward decays ($\cos \theta^{\ast} > 0$) and
\numb~the number of backward decays ($\cos \theta^{\ast} < 0$). 
The Collins-Soper frame is used because it minimises the impact of the transverse momentum of the incoming quarks on the identification of forward and backward decays. 

In this paper the asymmetry of the angular distribution of muons in $Z \rightarrow \mu^{+}\mu^{-}$ decays\footnote{In the following $Z$ is used to denote the \zy~contributions.} 
is measured using proton proton collision data collected by the LHCb experiment at centre-of-mass energies of $\sqs = 7$ and 8 \tev,
corresponding to an integrated luminosity of $1\fb^{-1}$ and $2\fb^{-1}$ respectively.
The asymmetry as a function of the dimuon invariant mass is used to determine \ssqtwef.

Comparisons of the determinations of the weak mixing angle from processes with different initial and final state fermions provide a test of the universality of the fermion to $Z$ couplings.
The most accurate measurement of \ssqtwef at the LEP experiments was obtained from the forward-backward asymmetry in $b$ quark final states~\cite{ALEPH:2005ab},
and at the SLD experiment by measuring the left-right asymmetry with polarised electrons~\cite{Abe:435742}. 
Determinations of \ssqtwef have also been obtained in hadronic
production processes with leptonic final states at the CDF and D0 experiments at the Tevatron~\cite{Aaltonen:2014loa,Abazov:2014jti} 
and ATLAS and CMS experiments at the LHC~\cite{Aad:2015uau,Chatrchyan:2011ya}.

Measurements of \afb~can be related to \ssqtwef when the momentum direction of the initial quark and antiquark are known.
At the LHC the momentum direction of the initial-state quark is not known, diluting the ability to determine \ssqtwef from \afb. 
However, since at LHC the dominant production process is $u \bar{u}, d \bar{d} \rightarrow Z$, the main contribution originates from a collision of a valence
quark with high momentum and a sea antiquark with lower momentum,
and so the $Z$ boson tends to be boosted along the direction of the quark.
This is particularly true in the forward region where the $Z$ boson has large longitudinal momentum.
Consequently, the sensitivity of \afb~to \ssqtwef is greater at large rapidities of the $Z$ boson.
Using simulated samples, it is found that
in the LHCb acceptance the assignment of forward and backward decays is correct in $90\%$ of the time.

The layout of this paper is as follows.
Section~\ref{sec:Detector} describes the LHCb detector and the data samples used in the analysis.
The candidate selection and background determination are described in Sec.~\ref{sec:Selection}.
In Sec.~\ref{sec:afb} the \afb~measurements are presented and 
in Sec.~\ref{sec:ssqtw} the measurements are compared to next-to-leading order (NLO) theoretical predictions within the same kinematic region, 
and a value of \ssqtwef is determined.

\section{Detector and datasets}
\label{sec:Detector}
The \lhcb detector~\cite{Alves:2008zz, Aaij:2014jba} is a single-arm forward
spectrometer covering the \mbox{pseudorapidity} range $2<\eta <5$.
The detector includes a high-precision tracking system
consisting of a silicon-strip vertex detector surrounding the $pp$
interaction region, a large-area silicon-strip detector located
upstream of a dipole magnet with a bending power of about
$4{\rm\,Tm}$, and three stations of silicon-strip detectors and straw
drift tubes placed downstream of the magnet.
The magnet polarity can be reversed, so that detector-induced asymmetries can be studied and corrected for in the analyses.
The tracking system provides a measurement of momentum, \ptot, of charged particles with
a relative uncertainty that varies from 0.5\% at low momentum to 1.0\% at 200\gev.\footnote{Units where the speed of light is set to unity are used throughout this paper.}
The minimum distance of a track to a primary vertex, the impact parameter, is measured with a resolution of $(15+29/\pt)\mum$,
where \pt is the component of the momentum transverse to the beam, in\,\gev.
Different types of charged hadrons are distinguished using information
from two ring-imaging Cherenkov detectors.
Photons, electrons and hadrons are identified by a calorimeter system consisting of
scintillating-pad and preshower detectors, an electromagnetic
calorimeter and a hadronic calorimeter. Muons are identified by a
system composed of alternating layers of iron and multiwire
proportional chambers.
The online event selection is performed by a trigger,
that consists of a hardware stage, based on information from the calorimeter and muon
systems, followed by a software stage, which applies a full event
reconstruction.
For this analysis, candidates are triggered by at least one muon with momentum larger than 10\gev.

Simulated samples are used to estimate the shapes of the invariant mass distributions for the simulated signal sample and some of the background sources.
The signal sample is also used to correct the data for reconstruction and detector effects. 
In the simulation, $pp$ collisions are generated using \pythia~8~\cite{Sjostrand:2006za, Sjostrand:2007gs} with a specific \lhcb configuration~\cite{LHCb-PROC-2010-056}.
Decays of hadronic particles are described by \evtgen~\cite{Lange:2001uf}, in which final-state radiation (FSR) is generated using \photos~\cite{Golonka:2005pn}. 
The interaction of the generated particles with the detector, and its response, are implemented using the \geant toolkit~\cite{Allison:2006ve, *Agostinelli:2002hh}, as described in Ref.~\cite{LHCb-PROC-2011-006}.

To simulate \zymm~decays with different values of \ssqtwef, the next-to-leading order generator \powhegbox~\cite{Alioli:2008gx}, interfaced to \pythia for the parton showering, is used.
Additional simulated samples are generated without parton showering using \powhegbox, \herwig\cite{Bahr:2008pv} and \fewz\cite{Gavin:2010az} and are used to evaluate 
theoretical uncertainties.
Predictions are also obtained using \fewz at NLO and are used to provide an alternative calculation of \afb~to compare to those computed by the \powhegbox~generator.
The same parton density function (PDF) is used for both generators.

\section{Event selection}
\label{sec:Selection}

Dimuon candidates, consisting of two oppositely charged muons, are selected using the same criteria as in Ref.~\cite{LHCb-PAPER-2015-001}, but with an extended mass range. 
The two muons must be within $2.0 < \eta < 4.5$,  have good quality track fits, a transverse momentum greater than $20\gev$ and must combine to an invariant mass within $60 < m_{\mu\mu} < 160\gev$.
These requirements define the kinematic region of this measurement.

The purity of the candidate sample is determined by estimating the contribution from background sources using a combination of simulation and data-driven techniques, and is found to be greater than $99\%$. 
The total yield, reconstructed dimuon invariant mass and \afb~are determined for each source of background.

\begin{figure}[tb]
  \begin{center}
    \includegraphics[width=0.495\linewidth]{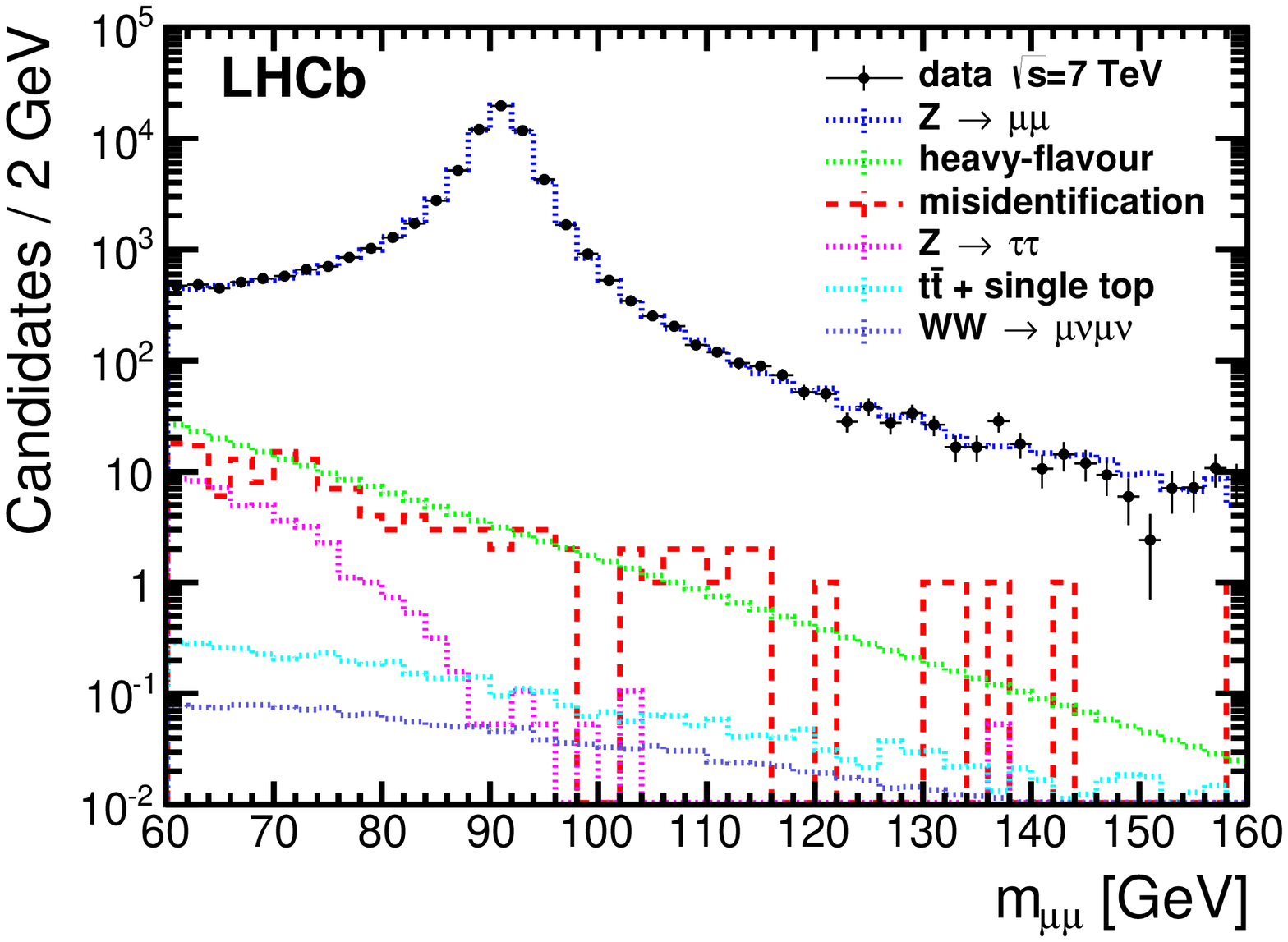} 
    \includegraphics[width=0.495\linewidth]{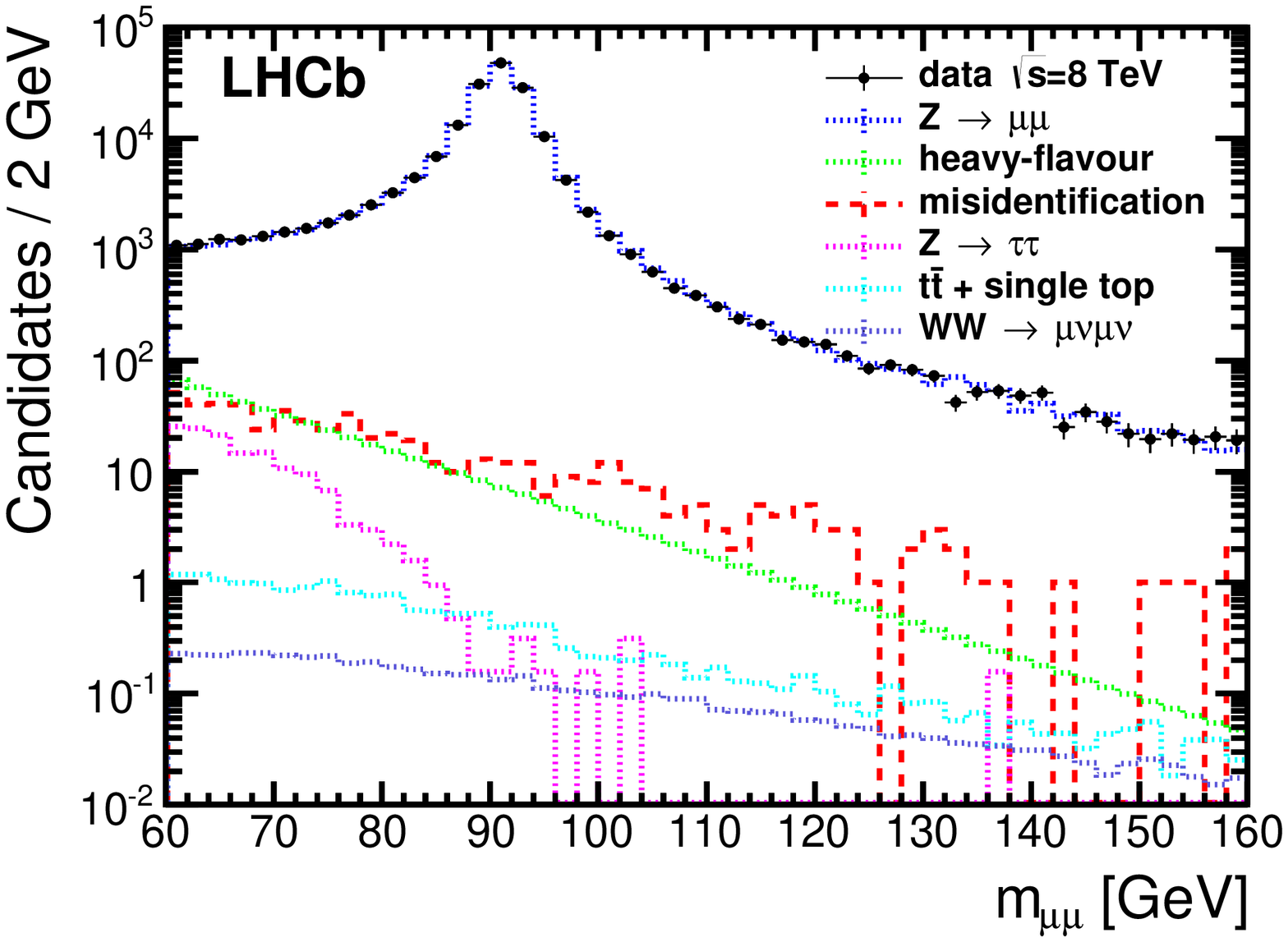}
    \vspace*{-1.0cm}
  \end{center}
  \caption{Dimuon invariant mass distribution for data, simulated signal and background processes for (left) {\sqs $=$ 7 TeV} and (right) {\sqs $=$ 8 TeV}.}
  \label{fig:invmass}
\end{figure}

The largest background contributions come from semileptonic heavy-flavour decays and events containing misidentified hadrons, where hadrons punching through the calorimeters to the muon stations are identified as muons, or hadrons have decayed in flight. 
Both contributions are estimated using data-driven techniques.
Two heavy-flavour enriched samples are selected by widening the mass window and requiring evidence that (i) the muons are produced away from the primary vertex, or (ii) that the muons are surrounded by hadronic activity.
These two samples are combined
to estimate both the shape of the reconstructed $m_{\mu\mu}$ distribution and the total number of events for the heavy-flavour background source.
The misidentified hadron contribution is estimated by using a sample of same-sign muon events. %selecting events which pass the selection, but which contain muons of the same charge.
The \ztautau, \ttbar, single top and \ww background sources are estimated using simulation.
The total background contribution is largest at low invariant mass.
The charge asymmetry of each background component is consistent with zero over the whole mass range.
The distribution of the dimuon invariant mass is shown for data and all background sources in Fig.~\ref{fig:invmass}.

\section{Forward-backward asymmetry measurements}
\label{sec:afb}

The forward-backward asymmetry is calculated from the selected dimuon candidates. 
Corrections are applied to account for efficiencies, 
biases in the reconstructed momenta of the muons
and differences in resolution between simulation and data.
Previous studies~\cite{LHCb-PAPER-2015-001, LHCb-PAPER-2014-033, LHCb-PAPER-2012-008} have observed a dependence of trigger, track reconstruction and muon identification efficiencies on muon pseudorapidity.
To account for this dependence, correction factors are evaluated from data using a tag-and-probe method~\cite{LHCb-PAPER-2015-001} and applied to the measured forward-backward asymmetry.

The momentum measurement of a muon is sensitive both to uncertainties in the detector alignment and the magnetic field scale. 
The magnetic field scale has been calibrated using dimuon and other resonances at low mass and is known to a precision of $0.04\%$~\cite{Aaij:2014jba}.
Low-mass resonances have also been used as input to the detector alignment, leading to a well-understood momentum calibration for low-momentum tracks~\cite{Amoraal:2012qn,LHCb-PAPER-2012-048,Aaij:2014jba}.
However, studies for the analysis presented here, have revealed a small, but appreciable, dependence of the position of the $Z$ resonance peak on muon kinematics. 
This effect can be attributed to residual detector misalignment. 
The corresponding muon curvature bias can be effectively parameterised in bins of the azimuthal angle of the muon about the beam axis. 
The parameters are determined using the difference between the $Z$ mass peak in data and simulation. % as input. 
The procedure is applied 
separately to data collected at $\sqs = 7$ and 8 \tev, and for the two magnet polarities.
The results are consistent with those presented in Ref.~\cite{LHCb-PAPER-2014-033} in which a slightly different method was used.

To compare with theory predictions, the data are unfolded for acceptance and resolution effects.
A Bayesian unfolding technique~\cite{D'Agostini:1994zf} is applied to the reconstructed dimuon invariant mass distribution~\cite{Adye:2011gm}. 
The unfolding algorithm is trained on simulation by comparing the generated invariant mass to that after reconstruction. 
The simulation is corrected to have the same $m_{\mu\mu}$ resolution as observed in data. 
Finally, the data are corrected for background by subtracting the distribution for each background source determined as described in Sec.~\ref{sec:Selection}.
No correction is applied to the measured values of \afb~to account for the dilution
due to imperfect knowledge of the initial quark direction, or to remove FSR effects. 
Instead, they are compared to predictions made within the same kinematic region and including FSR, as described in Sec.~\ref{sec:ssqtw}.

The following systematic uncertainties are considered when determining \afb.
The systematic uncertainty associated with the curvature correction is evaluated by varying these parameters within their uncertainty.
The uncertainties on the calibration factors are dependent on the sample size, and are therefore larger for the \sqs$= 7$\tev dataset.
This is the largest source of systematic uncertainty.
An uncertainty of $\pm 0.04\%$ is used for the momentum scale, determined from measurements of the magnetic field~\cite{Aaij:2014jba}.
The bias in the unfolding procedure is determined from simulation by comparing unfolded samples with the generated true $m_{\mu\mu}$ distribution.
%The uncertainty includes the dependence on the number of iterations used in the training of the unfolding algorithm.
An additional uncertainty to account for the dependence on the number of iterations used in the training of the unfolding algorithm is determined.
This variation has a larger effect in regions where fewer events are simulated.
The asymmetry of each background source does not vary significantly over the invariant mass range.
An uncertainty of $10\%$ is assigned to the background asymmetry, that covers the fluctuations observed in \afb~for each background source.
The effect of the uncertainties in the efficiency corrections applied to the data is found to be negligible.
The systematic uncertainties are determined separately for each bin of invariant mass and for both datasets. 
Their average values are summarized in Table~\ref{tab:afb_syst}.

\begin{table}[t]
\caption{Weighted average of the absolute systematic uncertainties for \afb, for different sources, given separately for \sqs $=$ 7 and 8~TeV.}
  \begin{center}\begin{tabular}{c c c}
    \hline
    Source of uncertainty                        & $\sqs = 7\tev$   & $\sqs = 8\tev$ \\ 
    \hline
    curvature/momentum scale          & 0.0102 & 0.0050 \\
    data/simulation mass resolution        & 0.0032 & 0.0025 \\
    unfolding parameter               & 0.0033 & 0.0009 \\
    unfolding bias                    & 0.0025 & 0.0025 \\ 

    \hline
  \end{tabular}\end{center}

\label{tab:afb_syst}
\end{table}

\begin{figure}[tb]
  \begin{center}
    \includegraphics[width=0.49\linewidth]{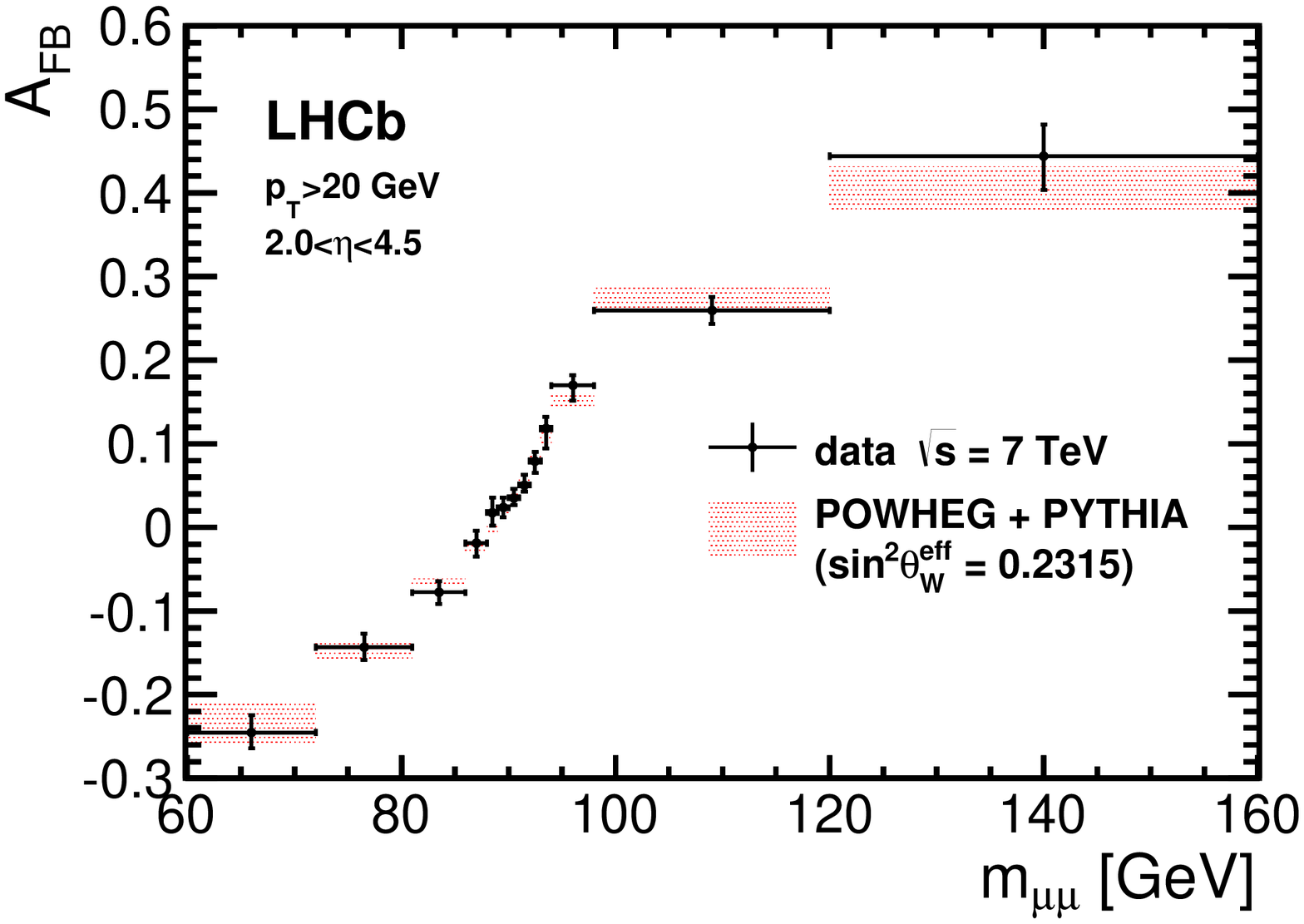}
    \includegraphics[width=0.49\linewidth]{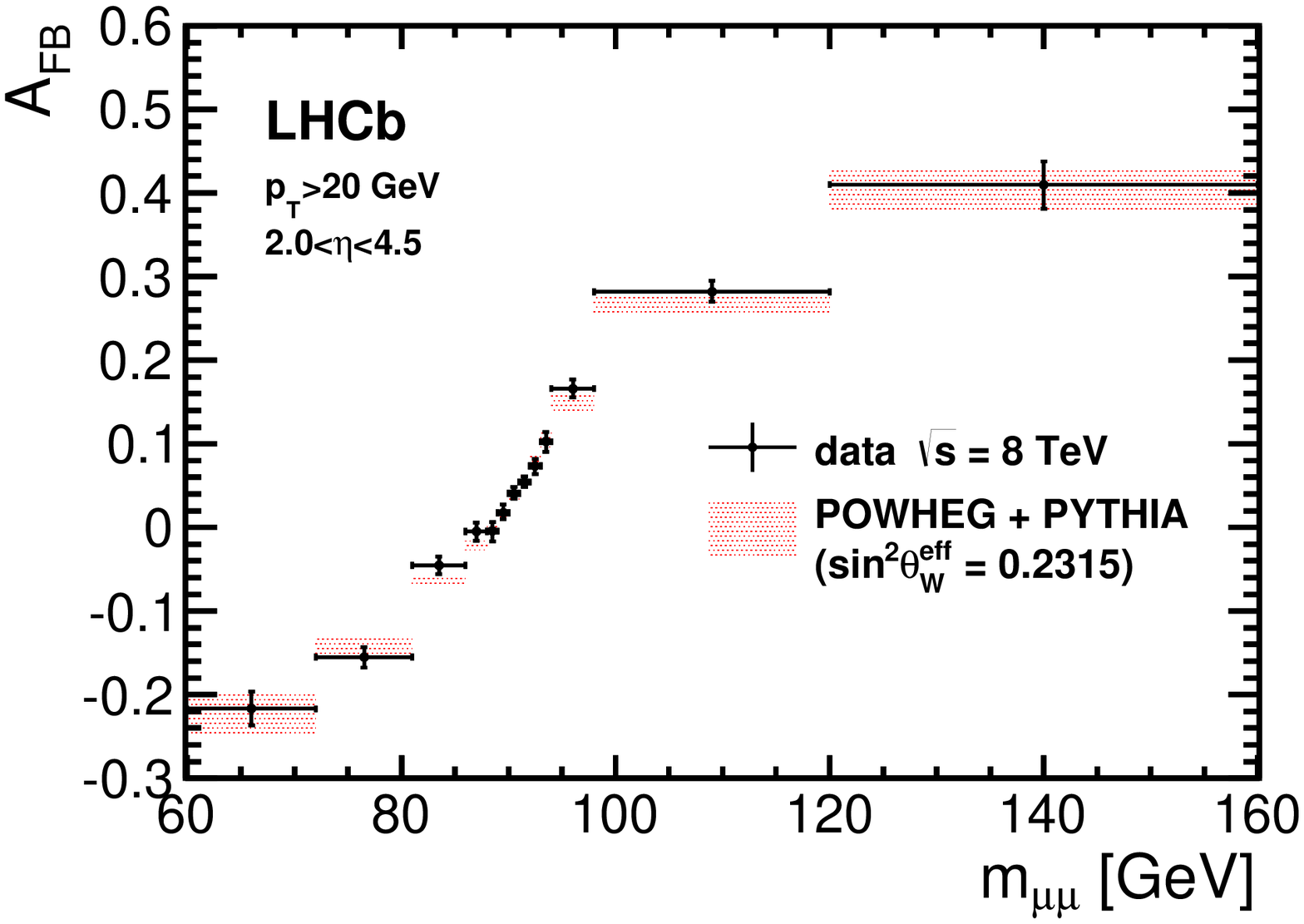}
    \vspace*{-0.4cm}
  \end{center}
  \caption{The measurements of \afb~as a function of the dimuon invariant mass for data compared to SM predictions for (left) {$\sqs = 7$~TeV} and (right) {$\sqs = 8$~TeV}. 
    The SM predictions are calculated using \powheg interfaced with \pythia for parton showering with the world average value for $\ssqtwef=0.2315$~\cite{PDG2014}. 
    The data include both statistical and systematic uncertainties, and the SM predictions include the theoretical uncertainties described in Sec.~\ref{sec:ssqtw}. 
  }
  \label{fig:afberr}
\end{figure}

The resulting measurements of \afb~for
$\sqs =$ 7 and  8 \tev data as a function of $m_{\mu\mu}$ are shown in Fig.~\ref{fig:afberr} 
and tabulated in Tables~\ref{tab:afb_results1} and~\ref{tab:afb_results2}. 

\begin{table}[t]

  \caption{Values for \afb~with the statistical and positive and negative systematic uncertainties for {\sqs $= 7$~TeV} data.
    The theoretical uncertainties presented in this table, corresponding to the PDF, scale and FSR uncertainties described in Sec.~\ref{sec:ssqtw}, affect only the predictions of \afb~and the \ssqtwef determination, and do not apply to the uncertainties on the measured \afb.}
  \begin{center} \begin{tabular}{c c c c c c}
    \hline
    $m_{\mu\mu}(\rm{GeV})$ & \afb       & stat.      & syst. $+$  & syst. $-$  & theoretical \\ 
    \hline                           
                                     
    $60 -72  $ & $-0.248             $ & $0.018   $ &  $0.011$   & $0.006$   & $0.025   $  \\ 
    $72 -81  $ & $-0.144             $ & $0.015   $ &  $0.006$   & $0.003$   & $0.011   $  \\ 
    $81 -86  $ & $-0.078             $ & $0.013   $ &  $0.005$   & $0.005$   & $0.005   $  \\ 
    $86 -88  $ & $-0.017             $ & $0.013   $ &  $0.007$   & $0.009$   & $0.005   $  \\ 
    $88 -89  $ & $\phantom{-}0.016   $ & $0.013   $ &  $0.012$   & $0.008$   & $0.005   $  \\ 
    $89 -90  $ & $\phantom{-}0.023   $ & $0.010   $ &  $0.006$   & $0.006$   & $0.005   $  \\ 
    $90 -91  $ & $\phantom{-}0.033   $ & $0.008   $ &  $0.007$   & $0.004$   & $0.005   $  \\ 
    $91 -92  $ & $\phantom{-}0.047   $ & $0.008   $ &  $0.009$   & $0.002$   & $0.005   $  \\ 
    $92 -93  $ & $\phantom{-}0.082   $ & $0.010   $ &  $0.004$   & $0.010$   & $0.006   $  \\ 
    $93 -94  $ & $\phantom{-}0.127   $ & $0.014   $ &  $0.004$   & $0.016$   & $0.009   $  \\ 
    $94 -98  $ & $\phantom{-}0.175   $ & $0.012   $ &  $0.003$   & $0.014$   & $0.009   $  \\ 
    $\phantom{1}98 -120 $ & $\phantom{-}0.259   $ & $0.015   $ &  $0.007$   & $0.006$   & $0.014   $  \\ 
    $120 -160 $ & $\phantom{-}0.451   $ & $0.037   $ &  $0.004$   & $0.017$   & $0.026   $  \\
    
    \hline
  \end{tabular}\end{center}

\label{tab:afb_results1}
\end{table}

\begin{table}[t]
  \caption{Values for \afb~with the statistical and positive and negative systematic uncertainties for {\sqs $= 8$~TeV} data.
    The theoretical uncertainties presented in this table, corresponding to the PDF, scale and FSR uncertainties described in Sec.~\ref{sec:ssqtw}, affect only the predictions of \afb~and the \ssqtwef determination, and do not apply to the uncertainties on the measured \afb.}
  \begin{center}\begin{tabular}{c c c c c c}
    \hline
    $m_{\mu\mu}$ $(\rm{GeV})$   & \afb     & stat.      & syst. $+$ & syst. $-$   & theoretical \\ 
    \hline
    $60 -72  $ & $-0.217  $            & $0.014   $ &  $0.015$  & $0.014$    & $0.025 $  \\ 
    $72 -81  $ & $-0.154  $            & $0.012   $ &  $0.004$  & $0.004$    & $0.011 $  \\ 
    $81 -86  $ & $-0.046  $            & $0.010   $ &  $0.003$  & $0.002$    & $0.005 $  \\ 
    $86 -88  $ & $-0.004  $            & $0.010   $ &  $0.003$  & $0.004$    & $0.005 $  \\ 
    $88 -89  $ & $-0.002  $            & $0.011   $ &  $0.003$  & $0.007$    & $0.005 $  \\ 
    $89 -90  $ & $\phantom{-}0.016   $ & $0.008   $ &  $0.006$  & $0.002$    & $0.005 $  \\ 
    $90 -91  $ & $\phantom{-}0.040   $ & $0.006   $ &  $0.005$  & $0.003$    & $0.005 $  \\ 
    $91 -92  $ & $\phantom{-}0.053   $ & $0.006   $ &  $0.004$  & $0.002$    & $0.005 $  \\ 
    $92 -93  $ & $\phantom{-}0.075   $ & $0.008   $ &  $0.004$  & $0.006$    & $0.006 $  \\ 
    $93 -94  $ & $\phantom{-}0.104   $ & $0.011   $ &  $0.003$  & $0.006$    & $0.009 $  \\ 
    $94 -98  $ & $\phantom{-}0.166   $ & $0.010   $ &  $0.005$  & $0.006$    & $0.009 $  \\ 
    $\phantom{1}98 -120 $ & $\phantom{-}0.280   $ & $0.012   $ &  $0.006$  & $0.002$    & $0.014 $  \\ 
    $120 -160 $ & $\phantom{-}0.412   $ & $0.027   $ &  $0.005$  & $0.009$    & $0.026 $  \\ 

    \hline
  \end{tabular}\end{center}
\label{tab:afb_results2}
\end{table}

\section{Determination of $\mathbf{ {\mathrm{\bf{sin}}}^{2}\theta_{\mathrm{\bf{W}}}^{\mathrm{\bf{eff}}} }$}
\label{sec:ssqtw}

The forward-backward asymmetry as a function of the dimuon invariant mass is compared with several sets of SM predictions generated with different values of \ssqtwef, denoted as \afbt.
The predictions are generated using \powhegbox~with \ssqtwef values ranging from 0.22 to 0.24 for $\sqs =$ 7 and 8 \tev, 
and the $Z$ boson mass ($M_Z$) and the electromagnetic coupling constant ($\alpha_{\text{EM}}$) fixed to the world average values~\cite{PDG2014}. 
The PDF set from NNPDF~\cite{Ball:2012cx}\footnote{NNPDF 2.3 QCD + QED NLO.}, with the strong coupling constant $\as(M_Z) = 0.118$, was used when generating the \afbt~samples. 

Theoretical uncertainties associated with the \afbt~distributions are taken into account when determining \ssqtwef. 
They arise from the underlying PDF, the choice of renormalisation and factorisation scales, the value of \as used, and the FSR calculation.
Each of these uncertainties, referred to collectively as theoretical uncertainties, are obtained from simulation.
The same uncertainty is assigned to \afbt~at both $\sqs = 7$ and $8\tev$.

To estimate the theoretical uncertainty of the PDF set, 
one hundred replica samples are produced, each with a unique PDF set provided by NNPDF~\cite{Ball:2013hta}. 
The value of \afbt~is calculated as a function of $m_{\mu\mu}$ for each of these replicas, and the corresponding $68\%$ confidence level interval determined.
The size of this uncertainty is larger than the difference observed using CT10~\cite{Lai:2010vv} as an alternative PDF parameterisation.

Uncertainty in the PDFs affects \afbt~in a way that is correlated across all dimuon invariant mass bins.
The same systematic uncertainty is applied for both collision energies and is therefore fully correlated for the two samples.

The uncertainty due to the choice of renormalisation and factorisation scales is studied by varying them by a factor of 0.5 and 2~\cite{Hamilton:2013fea}. 
The uncertainty in the \ssqtwef determination due to the uncertainty in $\alpha_s$ is estimated by studying the impact of a variation of $\pm 0.002$ when generating samples using \powhegbox. 
This covers the current uncertainty on $\alpha_s$~\cite{PDG2014}.
For both the $\alpha_s$ and scale uncertainties the final uncertainty is estimated by fitting a constant across the mass range to the maximum and minimum deviations in \afbt~to minimise the effect of statistical fluctuations in the samples.

The uncertainty due to the implementation of FSR is treated as a theoretical uncertainty.
It is obtained by comparing the \afbt~from three different generators, \fewz, {\textsc Herwig++} and \powhegbox+\pythia, before and after FSR.
To be consistent with the \powhegbox~sample, the \fewz generator is configured at NLO and electroweak corrections are not included.
The maximum and minimum difference is then determined and used to estimate the systematic uncertainty associated with FSR. 
The average size of the separate theoretical uncertainties is summarised in Table~\ref{tab:theo_syst}, and the combined uncertainties as a function of invariant mass are given in Tables~\ref{tab:afb_results1} and~\ref{tab:afb_results2}.

\begin{table}[tb]
    \caption{Weighted average of the absolute systematic uncertainties for \afbt, for the different sources of theoretical uncertainty. 
    The value quoted for the PDF uncertainty corresponds to the $68\%$ confidence range, while for the others the maximum and minimum shifts are given. 
    The correlations among the invariant mass bins are not taken into account.}
  \begin{center}
    \begin{tabular}{c c}
    \hline
    Uncertainty                    & average $\Delta |\afbt|$ \\ 
    \hline
    PDF                            & 0.0062  \\
    scale                          & 0.0040  \\
    $\alpha_s$                     & 0.0030  \\
    FSR                            & 0.0016  \\ 
    \hline
  \end{tabular}\end{center}
\label{tab:theo_syst}
\end{table}

The \afbt~shapes from \powhegbox~were cross-checked using the \fewz generator at the same value of \ssqtwef and the differences were found to be negligible. 

\begin{figure}[tb]
  \begin{center}
    \includegraphics[width=0.70\linewidth]{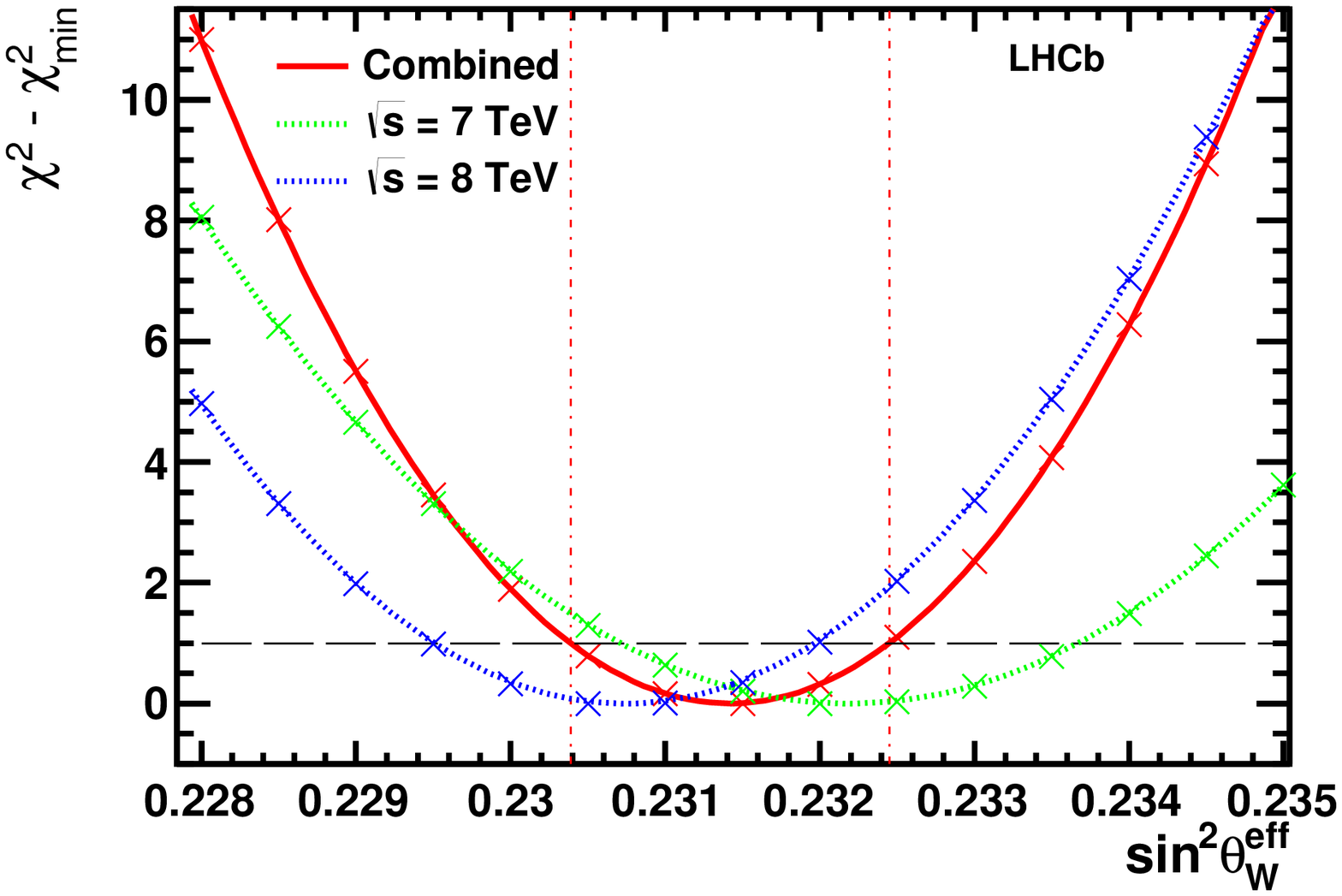}    
    \vspace*{-1.0cm}
  \end{center}
  \caption{Difference between the \chisq and the minimum \chisq obtained by comparing the final \afbm~measurements in data to \afbt~calculated using values of \ssqtwef ranging from 0.22 to 0.24, indicated by the crosses on the plot. 
    A quadratic fit is used to determine the minimum value for \ssqtwef and the corresponding uncertainty, and is shown for the different centre-of-mass energies and the combination.
The black dashed horizontal line corresponds to one unit of \chisq from the minimum and the intersecting \ssqtwef for the combination are indicated by the vertical red dashed lines.} 
  \label{fig:results_chi}
\end{figure}

The agreement between data and prediction is quantified by a \chisq value defined as
the square of the difference between the measured \afb~and \afbt~divided by the quadratic sum of the statistical, systematic and theoretical uncertainties, 
taking into account the correlations in the uncertainties between the mass bins. 
A quadratic function is fitted to the \chisq values of each set of fits as a function of \ssqtwef. 
The result is shown in Fig.~\ref{fig:results_chi}. 
The value of \ssqtwef at which \chisq takes its minimum is quoted as the final result for the \ssqtwef determination.
The interval in \ssqtwef corresponding to a variation of one unit in \chisq is quoted as the uncertainty.
The observed minimum values for the \chisqndf of the fit are 0.59 and 0.58, for the 7 and 8 \tev samples, respectively. 
The minima correspond to \finalresulta~and \finalresultb~respectively.
Results are cross-checked using a set of pseudoexperiments with the same statistics and background fractions as those in data. 
The values of \afb~from the pseudoexperiments are fitted to the prediction, and the spread of the measured \ssqtwef~values agrees with the uncertainties in the values of the $7$ and $8$~\tev samples.
A combination of these results, taking into account the correlation between systematic uncertainties for each centre-of-mass energy as well as the invariant mass bins, is obtained by calculating the full covariance matrix for the statistical, systematic and theoretical uncertainties. 
This yields 

$$\finalresultcombfull,$$ 
\noindent
where the first uncertainty is statistical, the second systematic and the third theoretical.

\begin{figure}[tb]
  \begin{center}
    \includegraphics[width=0.9\linewidth]{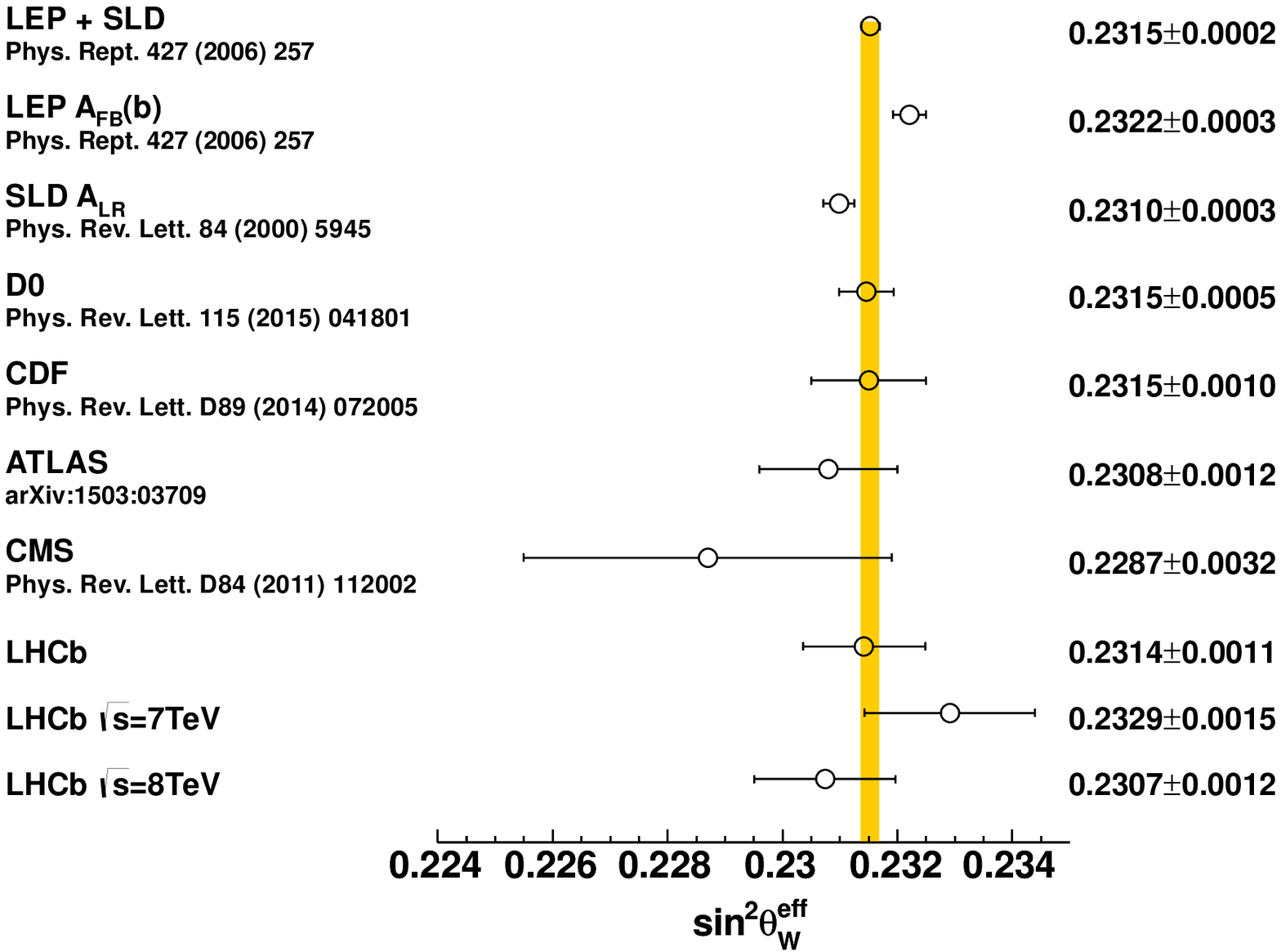}    
    \vspace*{-0.5cm}
  \end{center}
  \caption{A comparison of the \ssqtwef measurement at LHCb and other experiments.
  The combined LEP and SLD measurement is indicated by the vertical yellow band.}
  \label{fig:stw_comp}
\end{figure}

A comparison between the \ssqtwef result obtained here and those from other experiments is shown in Fig.~\ref{fig:stw_comp}.
The LHCb result agrees well with the world average and is one of the most precise measurements from hadron colliders.

\section{Conclusions}
\label{sec:Conclusion}
The forward-backward asymmetry for the process \qqZymm~as a function of the dimuon invariant mass is measured with the LHCb detector using proton proton collision data collected at centre-of-mass energies of $\sqs = 7$ and $8\tev$.
The measurements are performed in the Collins-Soper frame, using muons with $\pt > 20~$GeV and $2.0 < \eta < 4.5$ with a combined invariant mass $60 < m_{\mu\mu} < 160~$GeV.
The forward-backward asymmetry for each invariant mass bin is measured, together with the statistical and experimental uncertainties. 
The measurements at each centre-of-mass energy are used to determine a value for \ssqtwef,
by comparing to SM predictions that include FSR.
The best fit values obtained are \finalresulta~and \finalresultb~for the two samples at $\sqs = 7$ and $8\tev$ respectively. 
This leads to the combined result
 
$$\finalresultcombfull,$$
\noindent
where the first uncertainty is statistical, the second systematic and the third theoretical.
The measurement of \ssqtwef~presented here agrees with previous measurements.
The uncertainty from the PDF is the dominant theoretical uncertainty. Further high precision measurements at the LHC are expected to provide additional constraints in the forward region and reduce this uncertainty.
As the size of the data sample increases, it will become possible to perform a measurement of \afb~double-differentially in dimuon invariant mass and rapidity.
Such an approach will allow the analysis to take further advantage of the increased sensitivity of \afb~to \ssqtwef in the most forward region.

\section*{Acknowledgements}

\noindent We express our gratitude to our colleagues in the CERN
accelerator departments for the excellent performance of the LHC. We
thank the technical and administrative staff at the LHCb
institutes. We acknowledge support from CERN and from the national
agencies: CAPES, CNPq, FAPERJ and FINEP (Brazil); NSFC (China);
CNRS/IN2P3 (France); BMBF, DFG and MPG (Germany); INFN (Italy);
FOM and NWO (The Netherlands); MNiSW and NCN (Poland); MEN/IFA (Romania);
MinES and FANO (Russia); MinECo (Spain); SNSF and SER (Switzerland);
NASU (Ukraine); STFC (United Kingdom); NSF (USA).
We acknowledge the computing resources that are provided by CERN, IN2P3 (France), KIT and DESY (Germany), INFN (Italy), SURF (The Netherlands), PIC (Spain), GridPP (United Kingdom), RRCKI (Russia), CSCS (Switzerland), IFIN-HH (Romania), CBPF (Brazil), PL-GRID (Poland) and OSC (USA). We are indebted to the communities behind the multiple open
source software packages on which we depend. We are also thankful for the
computing resources and the access to software R\&D tools provided by Yandex LLC (Russia).
Individual groups or members have received support from AvH Foundation (Germany),
EPLANET, Marie Sk\l{}odowska-Curie Actions and ERC (European Union),
Conseil G\'{e}n\'{e}ral de Haute-Savoie, Labex ENIGMASS and OCEVU,
R\'{e}gion Auvergne (France), RFBR (Russia), XuntaGal and GENCAT (Spain), The Royal Society
and Royal Commission for the Exhibition of 1851 (United Kingdom).

\addcontentsline{toc}{section}{References}
\setboolean{inbibliography}{true}
\bibliographystyle{LHCb}
%\bibliography{main}
\bibliography{LHCb-PAPER-MODIFIED}

\newpage

%%%%%%%%%%%%%%%%%%%%%%%%%%%%%%%%%%%%%%%%%%
\centerline{\large\bf LHCb collaboration}
\begin{flushleft}
\small
R.~Aaij$^{38}$, 
B.~Adeva$^{37}$, 
M.~Adinolfi$^{46}$, 
A.~Affolder$^{52}$, 
Z.~Ajaltouni$^{5}$, 
S.~Akar$^{6}$, 
J.~Albrecht$^{9}$, 
F.~Alessio$^{38}$, 
M.~Alexander$^{51}$, 
S.~Ali$^{41}$
% ...

{\footnotesize \it
% ...
$ ^{5}$Clermont Universit\'{e}, Universit\'{e} Blaise Pascal, CNRS/IN2P3, LPC, Clermont-Ferrand, France\\
$ ^{6}$CPPM, Aix-Marseille Universit\'{e}, CNRS/IN2P3, Marseille, France\\
% ...
$ ^{9}$Fakult\"{a}t Physik, Technische Universit\"{a}t Dortmund, Dortmund, Germany\\
% ...
$ ^{37}$Universidad de Santiago de Compostela, Santiago de Compostela, Spain\\
$ ^{38}$European Organization for Nuclear Research (CERN), Geneva, Switzerland\\
% ...
$ ^{41}$Nikhef National Institute for Subatomic Physics, Amsterdam, The Netherlands\\
% ...
$ ^{46}$H.H. Wills Physics Laboratory, University of Bristol, Bristol, United Kingdom\\
% ...
$ ^{51}$School of Physics and Astronomy, University of Glasgow, Glasgow, United Kingdom\\
$ ^{52}$Oliver Lodge Laboratory, University of Liverpool, Liverpool, United Kingdom\\
% ...
}
\end{flushleft}
%%%%%%%%%%%%%%%%%%%%%%%%%%%%%%%%%%%%%%%%%%

\end{document}